\documentclass[conference,a4paper]{APSIPA2021}
\usepackage{multirow}
\usepackage[pdftex]{graphicx}
\usepackage{amsmath}
\usepackage[psamsfonts]{amssymb}
\usepackage{amsxtra}
\usepackage{threeparttable}

\usepackage{bm}
\usepackage{booktabs}

\makeatletter
\let\MYcaption\@makecaption
\makeatother
\usepackage[hang,small,bf]{caption}
\usepackage[subrefformat=parens]{subcaption}
\makeatletter
\let\@makecaption\MYcaption
\makeatother

\newcommand{\X}{\bm{X}}
\newcommand{\yp}{y^{\p}}
\newcommand{\yt}{y^{\t}}
\newcommand{\hatyt}{\hat{y}^{\t}}
\newcommand{\z}{\bm{z}}
\newcommand{\zp}{\z^{\p}}
\newcommand{\zt}{\z^{\t}}
\newcommand{\ztil}{\widetilde{\z}}
\newcommand{\zttil}{\ztil^{\t}}
\DeclareMathOperator*{\argmax}{argmax}

\newcommand{\atil}{\widetilde{\bm{a}}}
\newcommand{\btil}{\widetilde{\bm{b}}}
\newcommand{\vtil}{\widetilde{\bm{v}}}
\newcommand{\origin}{\mathbf{0}}
\newcommand{\origintil}{\widetilde{\origin}}
\newcommand{\mutil}{\widetilde{\bm{\mu}}}
\newcommand{\muyp}{\bm{\mu}_{\yp}^{\p}}
\newcommand{\muyttil}{\mutil_{\yt}^{\t}}
\newcommand{\mujtil}{\mutil_{j}^{\t}}
\newcommand{\mujprimetil}{\mutil_{j'}^{\t}}
\newcommand{\sig}{\bm{\sigma}}
\newcommand{\sigyp}{\sig_{\yp}^{\p}}
\newcommand{\sigyt}{\sig_{\yt}^{\t}}
\newcommand{\sigjt}{\sig_j^{\t}}
\newcommand{\sigjprimet}{\sig_{j'}^{\t}}
\newcommand{\Sig}{\bm{\Sigma}}

\newcommand{\xip}{\bm{\xi}^{\p}}
\newcommand{\xit}{\bm{\xi}^{\t}}
\newcommand{\xittil}{\widetilde{\bm{\xi}}^{\t}}
\newcommand{\etap}{\bm{\eta}^{\p}}
\newcommand{\etat}{\bm{\eta}^{\t}}
\newcommand{\wt}{\bm{w}^{\t}}
\newcommand{\epsp}{\bm{\varepsilon}^{\p}}
\newcommand{\epst}{\bm{\varepsilon}^{\t}}
\newcommand{\Dp}{D^{\p}}
\newcommand{\Dt}{D^{\t}}
\newcommand{\phip}{\phi^{\p}}
\newcommand{\phit}{\phi^{\t}}
\newcommand{\R}{\mathbb{R}}
\newcommand{\Rd}{\R^d}
\newcommand{\RD}{\R^D}
\newcommand{\Rdplusone}{\R^{d+1}}
\newcommand{\Rdd}{\R^{d \times d}}
\newcommand{\RDp}{\R^{\Dp}}
\newcommand{\RDt}{\R^{\Dt}}

\newcommand{\RDtDt}{\R^{\Dt \times \Dt}}
\renewcommand{\H}{\mathbb{H}}
\newcommand{\HKd}{\H_K^d}
\newcommand{\HKDt}{\H_K^{\Dt}}
\newcommand{\D}{\mathcal{D}}
\newcommand{\DHKd}{\D_{\HKd}}
\newcommand{\DHKDt}{\D_{\HKDt}}
\newcommand{\DKL}{\D_{\mathrm{KL}}}
\newcommand{\Dsame}{\D^{(\same)}}
\newcommand{\Ddiff}{\D^{(\diff)}}
\newcommand{\projK}{\mathrm{proj}^K}
\newcommand{\PTK}{\mathrm{PT}^K}
\newcommand{\expK}{\mathrm{exp}^K}
\newcommand{\logK}{\mathrm{log}^K}
\newcommand{\diag}{\mathrm{diag}}
\renewcommand{\L}{\mathcal{L}}
\newcommand{\ELBO}{\L (\theta, \phip, \phit; \X, \yp)}
\newcommand{\LCEp}{\L_{\CE}^{\p}}
\newcommand{\LCEt}{\L_{\CE}^{\t}}
\newcommand{\N}{\mathcal{N}}
\newcommand{\G}{\mathcal{G}}
\newcommand{\same}{\mathrm{same}}
\newcommand{\diff}{\mathrm{diff}}
\newcommand{\p}{(\mathrm{p})}
\renewcommand{\t}{(\mathrm{t})}
\newcommand{\CE}{\mathrm{CE}}

\begin{document}

\title{Hyperbolic Timbre Embedding\\for Musical Instrument Sound Synthesis\\Based on Variational Autoencoders}

\author{
\authorblockN{
Futa Nakashima\authorrefmark{1},
Tomohiko Nakamura\authorrefmark{1},
Norihiro Takamune\authorrefmark{1},\\
Satoru Fukayama\authorrefmark{2}, and
Hiroshi Saruwatari\authorrefmark{1}
}
\authorblockA{
\authorrefmark{1}The University of Tokyo, Tokyo, Japan}

\authorblockA{
\authorrefmark{2}National Institute of Advanced Industrial Science and Technology (AIST), Tokyo, Japan}
}

\maketitle
\thispagestyle{empty}

\begin{abstract}
    In this paper, we propose a musical instrument sound synthesis (MISS) method based on a variational autoencoder (VAE) that has a hierarchy-inducing latent space for timbre.
    VAE-based MISS methods embed an input signal into a low-dimensional latent representation that captures the characteristics of the input.
    Adequately manipulating this representation leads to sound morphing and timbre replacement.
    Although most VAE-based MISS methods seek a disentangled representation of pitch and timbre, how to capture an underlying structure in timbre remains an open problem.
    To address this problem, we focus on the fact that musical instruments can be hierarchically classified on the basis of their physical mechanisms.
    Motivated by this hierarchy, we propose a VAE-based MISS method by introducing a hyperbolic space for timbre.
    The hyperbolic space can represent treelike data more efficiently than the Euclidean space owing to its exponential growth property.
    Results of experiments show that the proposed method provides an efficient latent representation of timbre compared with the method using the Euclidean space.

\end{abstract}
\footnote[0]{This work was supported by JSPS KAKENHI Grant number JP19H01116.}
\section{Introduction}
    
    Musical instrument sound synthesis (MISS) is a technique of synthesizing a musical instrument sound under parametric control.
    The performance of MISS methods has been greatly improved by the introduction of deep generative models, e.g., normalizing flows, generative adversarial networks, and variational autoencoders (VAEs) \cite{Huzaifah2021HAIMchap}.
    
    One reason for using a VAE for MISS is its capability of editing input data through a low-dimensional latent space.
    A VAE embeds input data into the latent space and reconstructs the input from the embedded representation in the latent space (latent variable) in a Bayesian manner \cite{Kingma2014ICLR}.
    A latent variable can be seen as a parametric representation of the input.
    By manipulating the latent variable, we can generate a sound similar to but different from an input and replace the timbre of an input sound with another timbre \cite{Luo2019ISMIR, Luo2020ISMIR, Tanaka2021ICASSP, Caillon2021ArXiv}.
    
    \begin{figure}[t]
        \begin{center}
            \includegraphics[width=85mm]{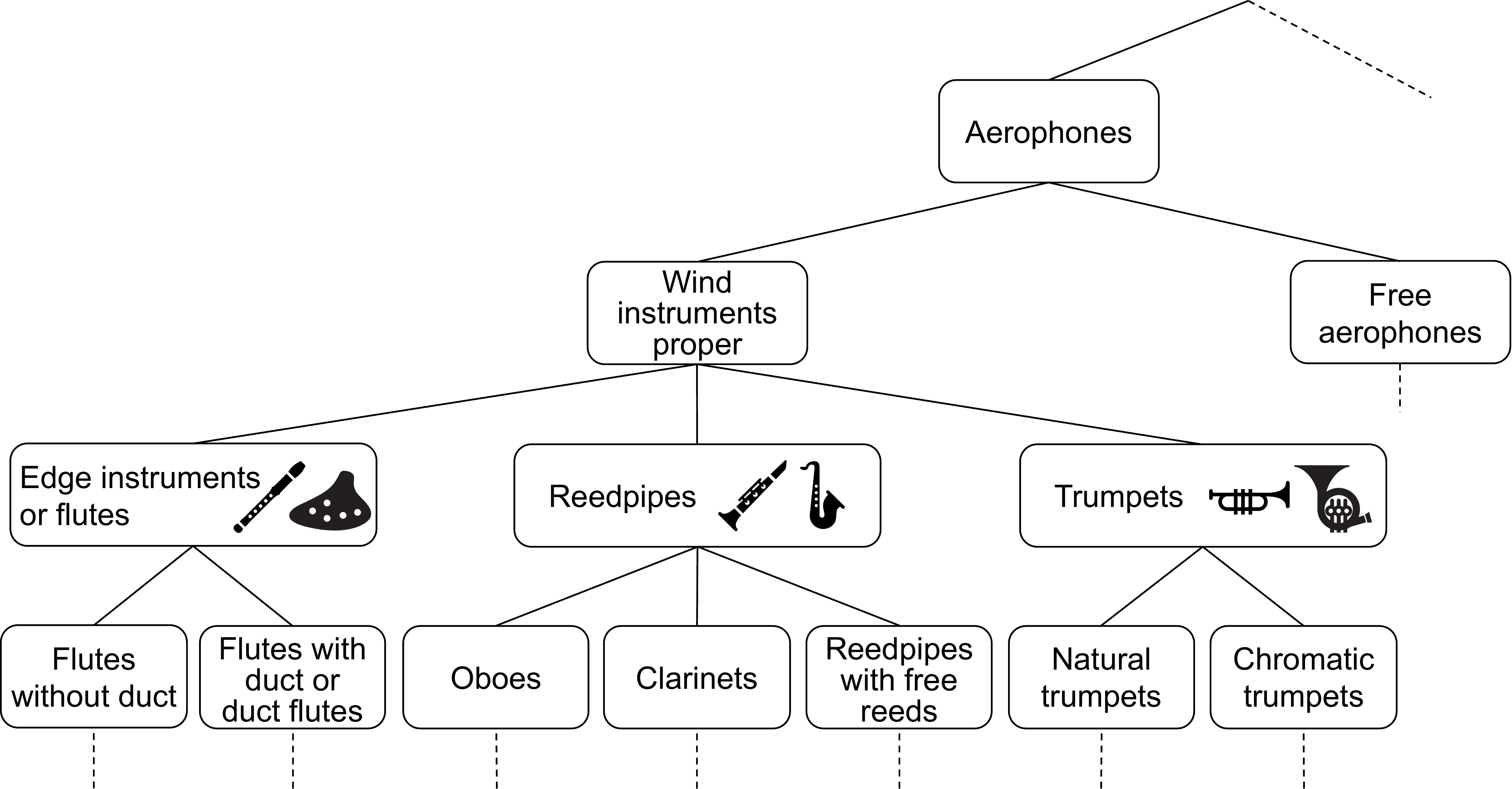}
        \end{center}
        \vspace*{-10pt}
        \caption{Excerpt of hierarchical classification system of musical instruments in \cite{Hornbostel1961GSJ}.}
        \label{fig:hornbostel}
        \vspace*{-3pt}
    \end{figure}
    
    In VAE-based MISS methods, the flexibility of editing a musical instrument sound mainly depends on how well the latent space captures the characteristics of musical instruments.
    In the Hornbostel--Sachs system for musical instrument classification, musical instruments are classified hierarchically \cite{Hornbostel1961GSJ} (see Fig.~\ref{fig:hornbostel}).
    The hierarchical system improves the accuracy of few-shot musical instrument recognition \cite{Garcia2021ISMIR}.
    
    We therefore utilize the hierarchical system for improving MISS.
    To capture hierarchical structures, we focus on a hyperbolic space, which is a non-Euclidean space with a constant negative curvature.
    In a Euclidean space, the distance from the origin increases linearly, whereas it increases exponentially in a hyperbolic space.
    Hierarchical data can be represented by a tree diagram and the number of nodes increases exponentially with tree depth, which coincides with the feature of the hyperbolic space.
    This implies that tree-structured data can be embedded more efficiently in a hyperbolic space than in a Euclidean space \cite{Sala2018ICML}.
    In fact, a hyperbolic space improves the performance of tasks on hierarchical data such as images, natural languages, and social graphs \cite{Peng2021TPAMI}.
    A hyperbolic space was incorporated into a VAE as a latent space, which has shown a superior performance for hierarchical data \cite{Nagano2019ICML}.
    We refer to the VAE with a hyperbolic latent space as the hyperbolic VAE.
    
    In this paper, we propose a VAE-based MISS method by introducing a hyperbolic space as a hierarchy-inducing latent space for timbre.
    The proposed method is based on a VAE-based MISS method proposed in \cite{Luo2019ISMIR} (we call it Luo's method), which assigns pitch and timbre to two distinct latent spaces.
    Owing to this separate modeling, we can use a hyperbolic space only for the latent space corresponding to timbre, whereas the latent space for pitch remains unchanged.
    The latent space of timbre is constructed by referring to the hyperbolic VAE.
    The hyperbolic VAE uses a probability distribution in a hyperbolic space named the pseudo-hyperbolic Gaussian distribution, which is reminiscent of a normal distribution in a Euclidean space.
    This distribution enables the computation of gradients with respect to its parameters.
    Thus, we can train the proposed model by a conventional gradient descent method.
    The proposed method is the first attempt to use a hyperbolic space for MISS to the best of our knowledge.
    By comparing the proposed method with Luo's one, we empirically confirm the effect of introducing hyperbolic spaces.

\section{Related Works} \label{sec:related}
    \subsection{VAEs} \label{sec:VAE}
        A VAE is one of the generative models based on deep neural networks (DNNs) \cite{Kingma2014ICLR}.
        It deals with a process of generating data $\X$ from a latent variable $\z \in \RD$ and represents $p(\X|\z)$ by a DNN named the decoder.
        The prior distribution of latent variable $p(\z)$ is assumed to be a standard multivariate normal distribution.
        
        The VAE seeks to find the set of decoder parameters $\theta$ that maximizes the log-likelihood $\log p(\X)$.
        Since it is difficult to calculate $\log p(\X)$ directly, we instead maximize a lower bound of $\log p(\X)$ [evidence lower bound (ELBO)] by introducing a variational posterior distribution $q(\z|\X)$:
        \begin{equation}
        \log p(\X)\geq
             \mathbb{E}_{\z \sim q(\z | \X)} \left[ \log p(\X | \z) \right] - \DKL  (q(\z | \X)  \| p(\z)), \label{eq:conventional_ELBO}
        \end{equation}
        where $\mathbb{E}_{\z \sim q}[\cdot]$ denotes an expectation over a random variable $\z$ following a probability distribution $q$, and $\DKL$ denotes the Kullback--Leibler divergence.
        The first term on the right-hand side of \eqref{eq:conventional_ELBO} induces the reconstructed data $\hat{\X}$ to be similar to $\X$, where $\hat{\X}$ is the output of the decoder.
        The second term induces $q(\z|\X)$ to be close to $p(\z)$.
        
        The key feature of the VAE is to estimate the parameters of $q(\z|\X)$ from $\X$ using a DNN called an encoder with the set of parameters $\phi$.
        Thus, the entire problem amounts to the problem of finding $\phi$ and $\theta$ that maximize ELBO \eqref{eq:conventional_ELBO}.

    \subsection{VAE-Based MISS Methods}
        Most VAE-based MISS methods focus on the disentanglement of pitch and timbre \cite{Luo2019ISMIR, Luo2020ISMIR, Tanaka2021ICASSP}.
        In \cite{Luo2019ISMIR}, Luo's method factorizes a latent space into a product space of two distinct latent spaces corresponding to pitch and timbre.
        Namely, it decomposes the latent variable $\z$ into two independent parts for pitch and timbre, $\zp \in \RDp$ and $\zt \in \RDt$, and it is assumed that $p(\z)=p(\zp)p(\zt)$.
        Note that $D=\Dp+\Dt$ and $\z=[(\zp)^\top, (\zt)^\top]^\top$, where $\top$ denotes the transposition of vectors.
        The parameters of the variational posterior distributions of $\zp$ and $\zt$ are separately estimated by two distinct encoders, whereas the decoder's input is $\z$, as in the usual VAE.
        This method has been extended to the problem that is blind to pitch and type of musical instrument by using contrastive learning \cite{Luo2020ISMIR}.
        As another approach to realizing better disentanglement, metric learning has been introduced in Luo's method \cite{Tanaka2021ICASSP}, which improves the disentanglement performance for unseen musical instruments.

\section{Hyperbolic Space}
    In this section, we briefly review the operations and the pseudo-hyperbolic Gaussian distribution in a hyperbolic space (see \cite{ Nagano2019ICML, Skopek2020ICLR} for details).
    A hyperbolic space is a non-Euclidean space whose curvature $K$ is constant and negative.
    Roughly speaking, the curvature $K$ is the quantity representing how a space is curved.
    A Euclidean space, which is used in a conventional VAE, has $K=0$, and a spherical space (e.g., hyperspheres) has $K>0$.
    Note that instead of the curvature $K$, the curvature radius $R=1/\sqrt{|K|}$ is sometimes used.
    In the following, $\widetilde{\cdot}$ means that a variable is defined in a hyperbolic space.

    \subsection{Lorentz Model}
        In \cite{Nagano2019ICML}, the Lorentz model is used as a hyperbolic space model.
        The advantage of the Lorentz model is that its basic operations can be described in closed form, as we will describe in Sections \ref{sec:exp_log_maps} and \ref{sec:parallel_transport}.
        
        The Lorentz model is defined by using a Lorentz inner product.
        For two $(d+1)$-dimensional real vectors, $\bm{a}=[a_1, \ldots, a_{d+1}]^\top$ and $ \bm{b}=[b_1,\ldots,b_{d+1}]^\top$,
        the Lorentz inner product is defined as 
        \begin{equation}
            \langle \bm{a}, \bm{b} \rangle_L = - a_1 b_1 + \sum_{i=2}^{d+1} a_i b_i .
        \end{equation}
        The $d$-dimensional Lorentz model with curvature $K$ is given by
        \begin{equation}
            \HKd = \left\{ \bm{a} \in \Rdplusone \Big| \langle \bm{a}, \bm{a} \rangle_L = \dfrac{1}{K},\ a_1 >0 \right\}.
        \end{equation}
        The point with minimal $a_1$, denoted by $\origintil = [1/\sqrt{-K}, 0, \dotsc , 0]^{\top}$, is called the origin.
        The distance of the shortest curve (geodesic) between two points $\atil, \btil \in \HKd$ is calculated as
        \begin{equation} \label{eq:hyperbolic distance}
            \DHKd (\atil, \btil) = \frac{1}{\sqrt{-K}} \cosh^{-1} (-K \langle \atil, \btil \rangle_L) .
        \end{equation}		

    \subsection{Exponential and Logarithm Maps} \label{sec:exp_log_maps}
        \begin{figure}[t]
            \begin{center}
                \includegraphics[width=70mm]{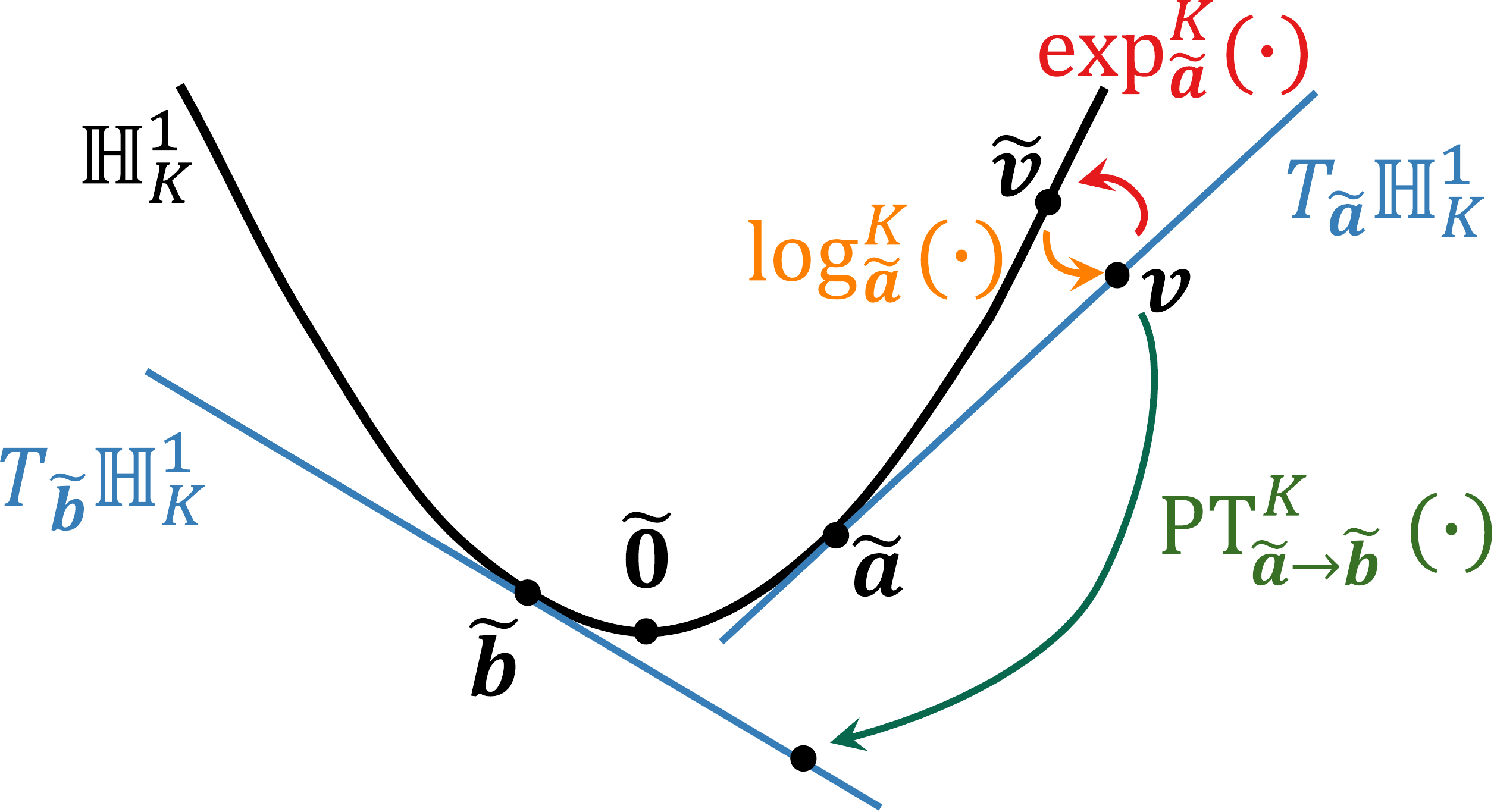}
            \end{center}
            \caption{Illustration of tangent space $T \HKd$ (blue lines), exponential map $\expK$ (red arrow), logarithm map $\logK$ (orange arrow), and parallel transport $\PTK$ (green arrow) in one-dimensional Lorentz model $\mathbb{H}_K^1$ (black curve).}
            \label{fig:hyperboloid}
            \vspace*{-3pt}
        \end{figure}
        
        The basic operations of $\HKd$ are defined via a tangent space (see Fig.~\ref{fig:hyperboloid}).
        For any point $\atil \in \HKd$ in the hyperbolic space (black curve), the tangent space (blue lines) $T_{\atil} \HKd \subset \Rdplusone$ is represented as the following $d$-dimensional vector space:
        \begin{equation}
            T_{\atil} \HKd = \left\{ \bm{u} \in \Rdplusone | \langle \bm{u}, \atil \rangle_L = 0 \right\}.
        \end{equation}
        
        The map from a tangent space to a hyperbolic space is called \emph{an exponential map}, and its inverse map is called \emph{a logarithm map}.
        The exponential map $\expK_{\atil}(\cdot)$ in the Lorentz model projects $ \bm{v}\in T_{\atil} \HKd$ into $\HKd$ as
        \begin{align}
            \expK_{\atil} (\bm{v}) &= \cosh (\sqrt{-K}  \lVert \bm{v} \rVert_L) \atil \notag \\
            &\quad +  \sinh (\sqrt{-K} \lVert \bm{v} \rVert_L) \frac{\bm{v}}{ \sqrt{-K} \lVert \bm{v} \rVert_L},
        \end{align}
        where the Lorentz norm $\lVert \cdot \rVert_L$ is given as $\lVert \bm{a} \rVert_L = \sqrt{\langle \bm{a}, \bm{a} \rangle_L}$.
        The logarithm map $\logK_{\atil} (\cdot)$ projects a point $\vtil \in \HKd$ on the hyperbolic space to $T_{\atil}\HKd$ as
        \begin{equation}
            \logK_{\atil} (\vtil) = \frac{\cosh^{-1} (K \langle \atil, \vtil \rangle_L) }{ \sinh (\cosh^{-1} (K \langle \atil, \vtil \rangle_L))} (\vtil - K \langle \atil, \vtil \rangle_L \atil). \label{eq:log_map}
        \end{equation}

    \subsection{Parallel Transport}\label{sec:parallel_transport}
        \emph{Parallel transport} in $\HKd$ is also defined via a tangent space, whereas it is defined as a vector addition in a Euclidean space.
        Parallel transport $\PTK_{\atil \rightarrow \btil} (\cdot)$ from a point $\atil$ to another point $\btil$ in $\HKd$ maps $\bm{v} \in T_{\atil} \HKd$ into another tangent space $T_{\btil} \HKd$ along a geodesic, while preserving the Lorentz inner product as
        \begin{equation}
            \langle \PTK_{\atil \rightarrow \btil} (\bm{v}), \PTK_{\atil \rightarrow \btil} (\bm{v'}) \rangle_L =  \langle \bm{v}, \bm{v'} \rangle_L
        \end{equation}
        for any $\bm{v}, \bm{v'} \in T_{\atil}\HKd$.
        It can be written analytically as
        \begin{equation}
            \PTK_{\atil \rightarrow \btil} (\bm{v}) = \bm{v} - \frac{K \langle \btil, \bm{v} \rangle_L }{1+K \langle \atil, \btil \rangle_L }(\atil + \btil).
        \end{equation}
        The inverse map can be defined by swapping the start and end points, i.e., $\PTK_{\btil \rightarrow \atil} (\cdot)$.
        For brevity, we denote the composition of $\PTK_{\origintil \rightarrow \atil}$ and $\expK_{\atil}$ as $\projK_{\atil}(\cdot)$:
        \begin{equation}
            \projK_{\atil}(\cdot) = \expK_{\atil} \circ \PTK_{\origintil \rightarrow \atil} (\cdot).
        \end{equation}
    		
    \subsection{Pseudo-hyperbolic Gaussian Distribution} \label{sec:pseudo}
        The pseudo-hyperbolic Gaussian distribution is a probability distribution in the hyperbolic space and reminiscent of a normal distribution in a Euclidean space \cite{Nagano2019ICML}.
        The pseudo-hyperbolic Gaussian distribution $\G (\ztil; \mutil, \Sig)$ of $\ztil \in \HKd$ has two parameters, mean $\mutil \in \HKd$ and covariance matrix $\Sig \in \Rdd$, similarly to a normal distribution.
        The random variable $\ztil \sim \G (\ztil; \mutil, \Sig)$ is obtained as 
        \begin{equation}
            \ztil = \projK_{\mutil} \left(\begin{bmatrix}0 \\ \bm{w}\end{bmatrix}\right),
            \label{eq:sampling}
        \end{equation}
        where $\bm{w} \in \Rd$ is a random variable that follows a $d$-dimensional normal distribution with zero mean $\origin_d$ and covariance $\Sig$, $\N (\bm{w}; \origin_d, \Sig)$.
        Considering the change of variables, we can calculate the log-probability density function of $\G (\ztil; \mutil, \Sig)$ as
        \begin{align}
            &\log p(\ztil ; \mutil, \Sig) \notag \\
            &= \log p(\bm{w}; \origin_d, \Sig) -\log \left| \det \frac{\mathrm{d} \ztil}{\mathrm{d} \bm{w}} \right| \nonumber \\
            &= \log p(\bm{w} ; \origin_d, \Sig) - (d-1) \log \left( \frac{ \sinh \left(\sqrt{-K} \lVert \bm{u} \rVert_L \right) }{\sqrt{-K} \lVert \bm{u} \rVert_L} \right),
        \end{align}
        where $\bm{u} = \logK_{\mutil} (\ztil)$.
        
        As mentioned above, the log-probability density function can be described analytically and the sampling from the pseudo-hyperbolic Gaussian distribution can be easily carried out by the use of \eqref{eq:sampling}.
        The gradients with respect to $\mutil$ and $\Sig$ can also be computed using a chain rule.
        Owing to these benefits, we can use the pseudo-hyperbolic Gaussian distribution as a hyperbolic-space counterpart of a normal distribution in a Euclidean space in the VAE framework \cite{Nagano2019ICML}.

\section{Hyperbolic Timbre Embeddings} \label{proposed}
    \begin{figure*}[t]
        \centering
        \includegraphics[width=170mm]{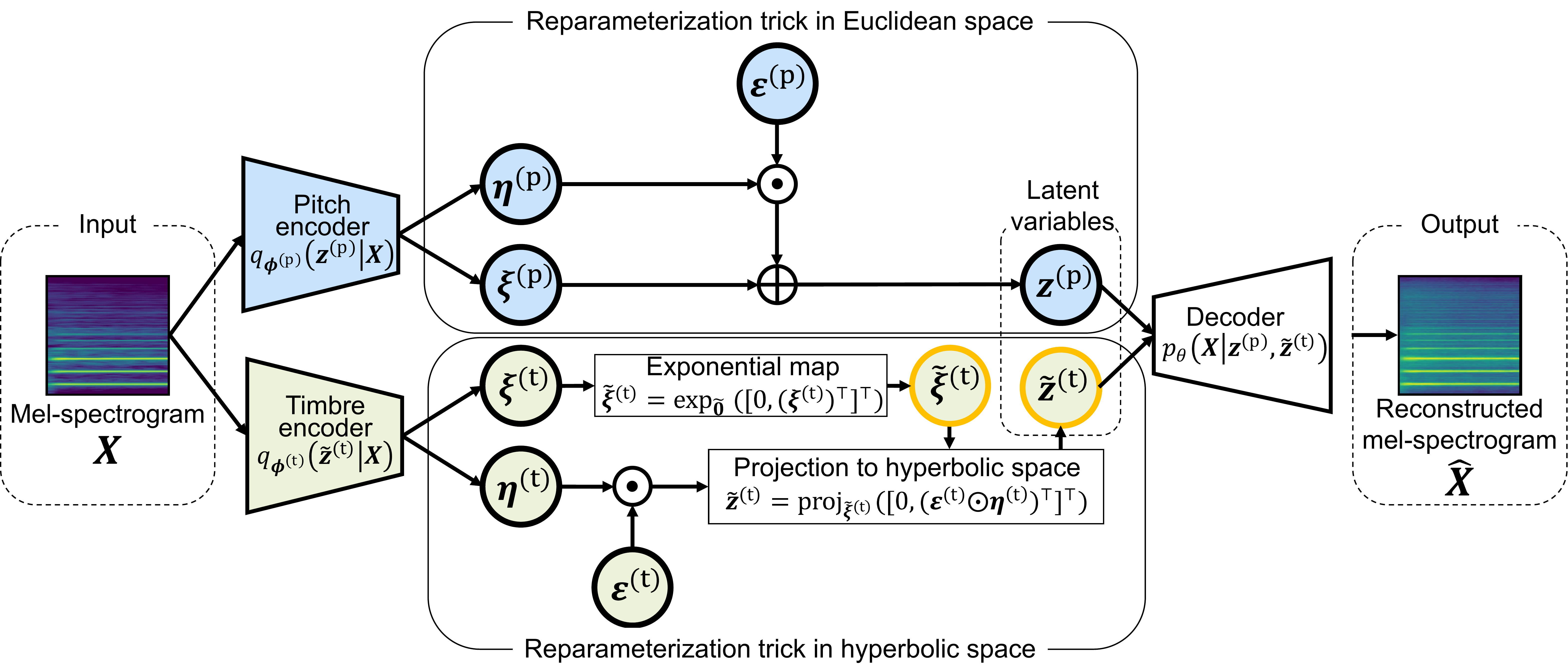}
        \caption{Overview of proposed VAE-based model and reparameterization tricks in Euclidean and hyperbolic spaces.}
        \label{fig:model}
    \end{figure*}
    In this section, we propose an extension of Luo's method by introducing a hierarchy-inducing latent space for timbre.
    For brevity, we reuse the identical notations of the random variables used in Section \ref{sec:related}.
    In the proposed method, we define the latent variable of timbre $\zttil$ in the hyperbolic space $\HKDt$ and introduce a pseudo-hyperbolic Gaussian distribution as a prior distribution of $\zttil$.
    The latent variable of pitch $\zp$ is defined in the same manner as in Luo's method.
    In the subsequent sections, we describe the generative model and training method of the proposed method.
    
    \subsection{Generative Model of Mel-spectrogram}
        The proposed model deals with a mel-spectrogram of a musical instrument sound as an input data $\X$.
        Additionally, the pitch label $\yp \in \mathcal{P}$ corresponding to the sound is given, where $\mathcal{P}$ is the set of possible pitches.
        
        The proposed model has three random variables other than $\X$: the timbre label $\yt \in \mathcal{T}$, the latent variable of pitch $\zp \in \RDp$, and the latent variable of timbre $\zttil \in \HKDt$, where $\mathcal{T}$ is the set of possible timbres.
        The timbre label $\yt$ follows the following categorical distribution:
        \begin{equation}
            \yt \sim \text{Cat} \left(\yt; \dfrac{1}{|\mathcal{T}|} \bm{1}_{|\mathcal{T}|}\right),
            \label{eq:categorical}
        \end{equation}
        where $|\cdot|$ returns the number of elements for a set and $\bm{1}_{d}$ is the $d$-dimensional vector whose elements are one.
        The latent variable of pitch $\zp$ follows a normal distribution with a mean and a diagonal covariance matrix conditioned on $\yp$:
        \begin{equation}
            \zp \sim \N (\zp; \muyp, (\diag(\sigyp))^2),
        \end{equation}
        where $\diag(\cdot)$ denotes a diagonal matrix whose diagonal elements are the input vector, $\muyp \in \RDp$ is the mean, and $\sigyp \in \RDp$ is a non-negative vector whose elements are standard deviations.
        $\zttil$ follows a pseudo-hyperbolic Gaussian distribution with mean $\muyttil \in \HKDt$ and covariance matrix $(\diag(\sigyt))^2 \in \RDtDt$ conditioned on $\yt$:
        \begin{equation}
            \zttil | \yt \sim \G (\zttil; \muyttil, (\diag(\sigyt))^2),
        \end{equation}
        where $\sigyt \in \RDp$ is a non-negative vector whose elements are standard deviations.
        The decoder represents the process of generating $\X$ from $\zp$ and $\zttil$, $p(\X| \zp, \zttil)$, by a DNN with the set of parameters $\theta$.
        
        In summary, the joint distribution of $\X, \zp, \zttil,$ and $\yt$ can be described as
        \begin{align}
            &p(\X, \zp, \zttil, \yt; \yp) \notag \\
            &=p(\X| \zp, \zttil)p(\zp; \yp)p(\zttil| \yt)p(\yt). \quad \label{eq:LL}
        \end{align}
        By marginalizing it out with respect to $\zp, \zttil,$ and $\yt$, we can compute the log-marginal distribution $\log p(\X; \yp)$.
        
    \subsection{Encoders of Pitch and Timbre}
        Similarly to a usual VAE (see Section \ref{sec:VAE}), the maximum likelihood estimation problem of $\log p(\X; \yp)$ becomes the problem of maximizing the ELBO of the log-likelihood $\log p(\X; \yp)$.
        As in \cite{Luo2019ISMIR}, we use a distinct DNN-based encoder for each of $\zp$ and $\zt$ (see Fig.~\ref{fig:model}).
        We call the encoders of $\zp$ and $\zt$ the pitch and timbre encoders, respectively.
        
        Let $\phip$ and $\phit$ be the DNN parameters of the pitch and timbre encoders, respectively.
        The pitch encoder transforms $\X$ into the parameters of the variational posterior distribution $q_{\phip}(\zp|\X)$, which is a $\Dp$-dimensional normal distribution with mean $\xip$ and diagonal covariance matrix $(\diag(\etap))^2$, i.e.,
        \begin{equation}
            q_{\phip}(\zp|\X) = \N(\zp;\xip,(\diag(\etap))^2).
        \end{equation}
        For numerical stability, the pitch encoder outputs not $\etap$ but its element-wise logarithm. 
        
        The timbre encoder converts $\X$ into the parameters of the variational posterior distribution $q_{\phit}(\zttil|\X)$, which is a $\Dt$-dimensional pseudo-hyperbolic Gaussian distribution with mean $\xittil$ and diagonal covariance matrix $(\diag(\etat))^2)$, i.e.,
        \begin{equation}
            q_{\phit}(\zttil|\X) = \G (\zttil; \xittil, (\diag(\etat))^2).
        \end{equation}
        As in the pitch encoder, it outputs the element-wise logarithm of $\etat$ instead of $\etat$.
        Unlike the pitch encoder, the timbre encoder does not directly output mean $\xittil$.
        To obtain $\xittil$, it outputs a $\Dt$-dimensional real vector $\xit$, treats it as a vector in the tangent space of $\HKDt$ around $\origintil$, and projects it onto $\HKDt$ using the exponential map of $\origintil$:
        \begin{equation}
            \xittil = \expK_{\origintil} \left( \begin{bmatrix}0 \\ \xit \end{bmatrix} \right).
        \end{equation}
            
    \subsection{Loss Function} \label{sec:loss}
        By using the variational posterior distributions $q_{\phit}(\zttil|\X)$ and $q_{\phit}(\zttil|\X)$, we can write the ELBO as
        \begin{align}
            & \ELBO \nonumber\\
            &= \mathbb{E}_{\substack{\zp \sim q_{\phip}(\zp | \X) \\ \zttil \sim q_{\phit}(\zttil | \X)}} \left[ \log p_{\theta}(\X | \zp, \zttil) \right] \nonumber\\
            &\quad - \DKL  (q_{\phip}(\zp | \X) \| p(\zp | \yp)) \nonumber\\
            &\quad - \mathbb{E}_{\yt \sim q(\yt | \X)} \left[ \DKL (q_{\phit}(\zttil | \X)  \| p(\zttil | \yt)) \right] \nonumber\\
            &\quad - \DKL  (q(\yt | \X) \| p(\yt)) \label{eq:ELBO},
        \end{align}
        where $q(\yt | \X)$ is the posterior distribution of $\yt$.
        As in Luo's method, we approximate $q(\yt | \X)$ as
        \begin{equation}
            q(\yt | \X)\simeq\mathbb{E}_{\zttil \sim q_{\phit}(\zttil | \X)}[p(\yt|\zttil)],
            \label{eq:approx}
        \end{equation}
        where $p(\yt|\zttil)$ is given as
        \begin{align}
            p(\yt=j|\zttil) &= \dfrac{p(\zttil|\yt=j)p(\yt=j)}{\sum_{j'\in\mathcal{T}}p(\zttil|\yt=j')p(\yt=j')} \nonumber \\
            &= \dfrac{\G (\zttil; \mujtil, (\diag(\sigjt))^2)}{\sum_{j'\in\mathcal{T}}\G (\zttil; \mujprimetil, (\diag(\sigjprimet))^2)}.
        \end{align}
        See Appendix A in \cite{Hsu2019ICLR} for details of this approximation.
        
        The loss function of Luo's method consists of two terms related to pitch and timbre classification in addition to the ELBO \cite{Luo2019ISMIR}.
        Following this strategy, we define a loss function as 
        \begin{align}
            -\ELBO + \LCEp + \LCEt.
            \label{eq:loss_func}
        \end{align}
        The term $\LCEp$ is a cross-entropy (CE) loss that involves the prediction of $\yp$ from $\zp$.
        For this term, a linear classifier is trained simultaneously.
        The classifier is implemented as a fully connected layer with $\Dp$ input units and $|\mathcal{P}|$ output units followed by a softmax nonlinearity.
        The CE loss is computed with the classifier's output and the ground-truth pitch label.
        The term $\LCEt$ is a CE loss that involves the prediction of $\yt$ from $\zttil$.
        This prediction problem is formulated as the problem of finding timbre label $j\in\mathcal{T}$ that maximizes $q(\yt|\X)$.
        In accordance with \eqref{eq:approx}, we obtain approximate values of $q(\yt=j|\X)$ for all $j$ and use them for the computation of $\LCEt$.

    \subsection{Reparameterization Tricks for Latent Variables in Euclidean and Hyperbolic Spaces}
        The proposed method seeks to find $\phip, \phit$, and $\theta$ that minimize the sum of loss function \eqref{eq:loss_func} over all training data.
        Owing to the complexity of the DNNs, it is difficult to analytically compute the expectations over $\zp$ and $\zttil$ in ELBO \eqref{eq:ELBO}.
        In a VAE framework, we frequently use an approximate computation method called the reparameterization trick proposed in \cite{Kingma2014ICLR}.
    
        For $\zp$, we can use the same reparameterization trick as in a usual VAE.
        We draw an instant $\epsp$ from a $\Dp$-dimensional multivariate standard normal distribution $\N( \epsp; \origin_{\Dp}, \bm{I}_{\Dp})$, where $\bm{I}_{\Dp}$ denotes the $\Dp \times \Dp$ identity matrix.
        The instant $\epsp$ is converted as
        \begin{equation}
            \xip + \epsp \odot \etap, \label{eq:rep_trick_euclid}
        \end{equation}
        where $\odot$ denotes the element-wise product.
        Since the linear transformation of a random variable following a normal distribution also follows a normal distribution, \eqref{eq:rep_trick_euclid} follows a $\Dp$-dimensional normal distribution with mean $\xip$ and diagonal covariance matrix $(\diag(\etap))^2$.
        
        For $\zttil$, the above reparameterization trick in the Euclidean space cannot be used because it is defined in the hyperbolic space.
        Since we use the pseudo-hyperbolic Gaussian distribution, we can use a reparameterization trick for it \cite{Nagano2019ICML}.
        We first draw $\epst \in \RDt$ from a standard normal distribution $\N(\epst; \origin_{\Dt}, \bm{I}_{\Dt})$.
        Then, we convert it as
        \begin{equation}
            \wt = \epst \odot \etat,
        \end{equation}
        which follows $\N(\wt; \origin_{\Dt}, (\diag(\etat))^2)$.
        Finally, we map $\wt$ into $\HKd$ via $T_{\xittil} \HKd$  as
        \begin{equation}
            \zttil = \projK_{\xittil}\left(\begin{bmatrix}0\\ \wt\end{bmatrix}\right).
            \label{eq:rep_trick_hyp}
        \end{equation}
        It results in the latent variable $\zttil$ sampled following $\G (\ztil; \xittil, (\diag(\etat))^2)$ in the way described in Section \ref{sec:pseudo}.
        Fig.~\ref{fig:model} shows the procedures of the above reparameterization tricks in the Euclidean and hyperbolic spaces.
        	
\section{Timbre Embedding Experiment}
    \subsection{Experimental Condition}
        We conducted experiments to examine the effect of using the hyperbolic space for timbre embedding.
        
        \subsubsection{Dataset}
            We used the Studio-On-Line dataset \cite{Ballet1999} as in \cite{Luo2019ISMIR}.
            It contains audio recordings of 1885 isolated notes played on 12 instruments (English horn, French horn, tenor trombone, trumpet, piano, violin, violoncello, saxophone, bassoon, clarinet, flute, and oboe) and 82 different pitches (A0 to F$\sharp$7).
            We used 1531 recordings for training, 177 for validation, and 177 for tests so that the musical instruments were uniformly included.
            The sampling frequency was 22.05~kHz.
            The first 500 ms of each recording was converted into a mel-spectrogram with 256 mel-frequency bins and 43 frames by  a short-time Fourier transform using a Hann window with a window size of 92~ms and a hop size of 11~ms.
            The center frequencies of the mel-filterbanks were from 27~Hz to 11~kHz with a uniform interval in the mel-frequency domain.
        	
        \subsubsection{Compared Methods}
            We compared the proposed method with Luo's method \cite{Luo2019ISMIR}, where the latent spaces of pitch and timbre are Euclidean spaces.
        	For evaluating the effects of the curvature and latent dimension, we varied $R$ as 1, 2, 5, 10, 20, 50, and 100 (corresponding to $K=-1, -1/4, -1/25, -1/100, -1/400, -1/2500,$ and $-1/10000$, respectively) and $\Dt$ as 2, 4, 8, and 16.
        	The latent space of pitch was set to $\Dp=16$, which was the same as that used in the experiments in \cite{Luo2019ISMIR}.

        \subsubsection{Network Architecture}
            \begin{figure}[t]
                \begin{minipage}[b]{\columnwidth}
                    \centering
                    \includegraphics[width=\columnwidth]{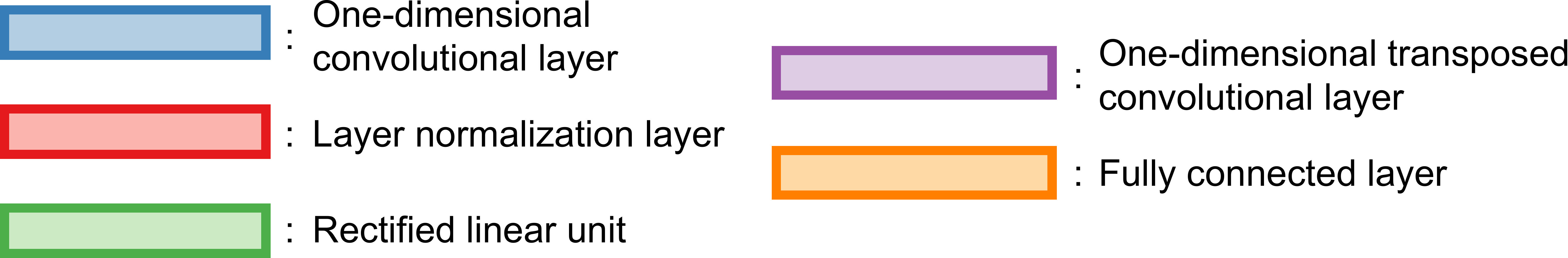}
                \end{minipage}
                
                \begin{minipage}[b]{0.64\columnwidth}
                    \centering
                    \includegraphics[width=53mm]{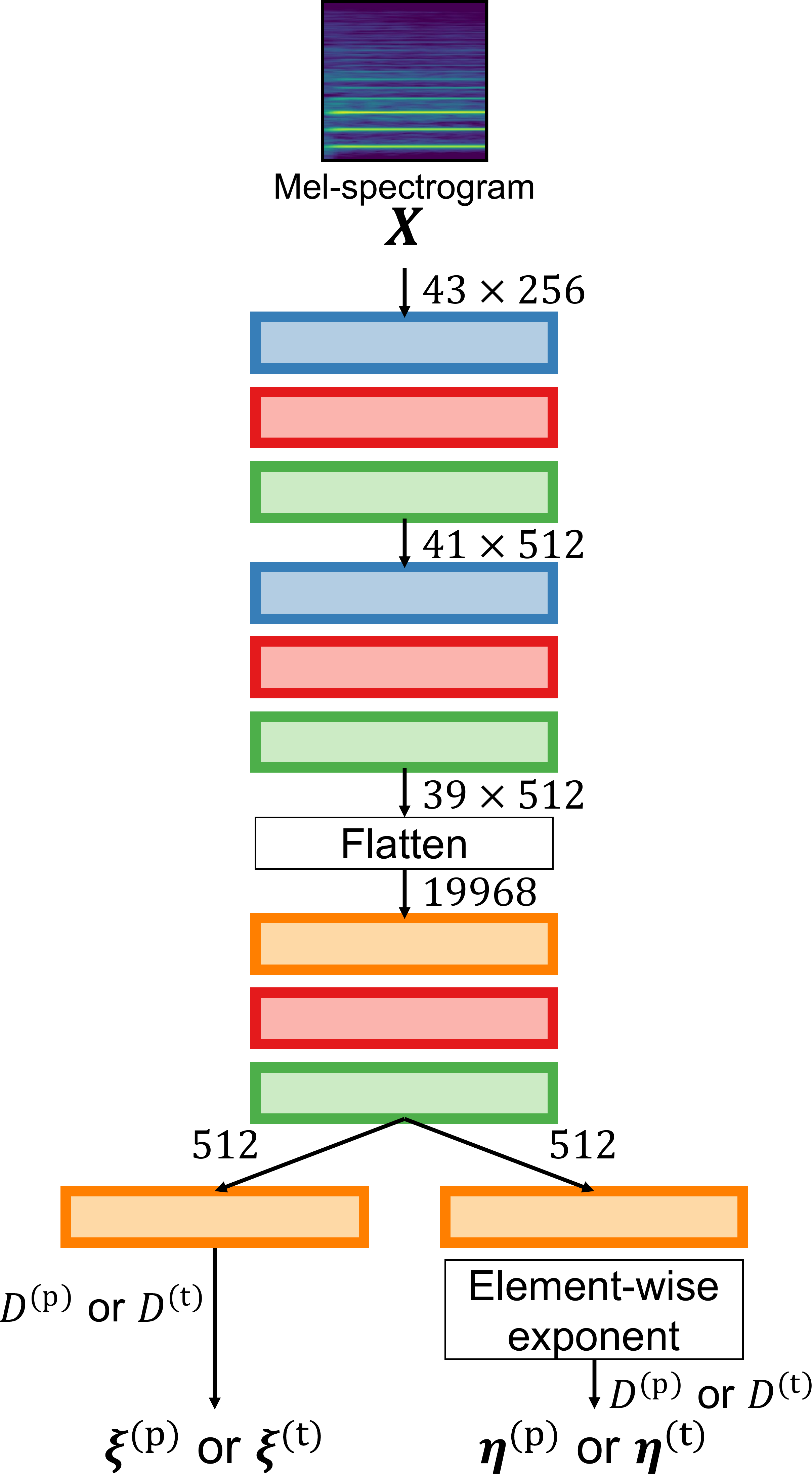}
                    \vspace{3pt}
                    \subcaption{Encoder}
                \end{minipage}
                \hspace{0.02\columnwidth}
                \begin{minipage}[b]{0.31\columnwidth}
                    \centering
                    \includegraphics[width=25mm]{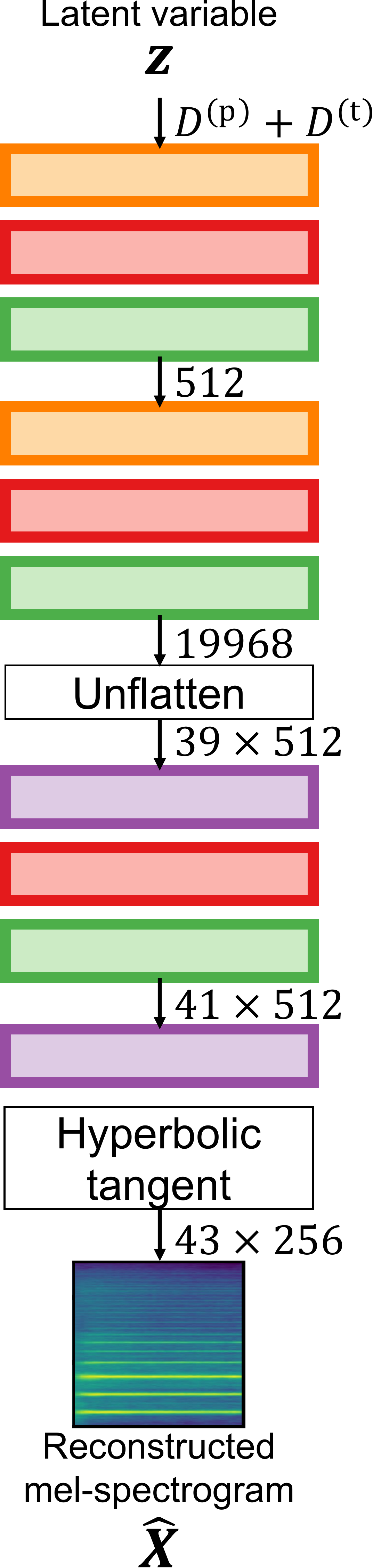}
                    \vspace{6pt}
                    \subcaption{Decoder}
                \end{minipage}
                \caption{Encoder and decoder architectures. Values beside arrows indicate dimensions of features.}
                
                \label{fig:network}
            \end{figure}
            
            Fig.~\ref{fig:network} shows the architectures of the encoder and decoder, where convolutional and transposed convolutional layers have a kernel size of $3$ and a stride of $1$.
            The pitch and timbre encoders had the same architecture.
            This structure is identical to that used in \cite{Luo2019ISMIR} except for using layer normalization layers instead of batch normalization layers.
            For fair comparison, the DNNs of encoders and decoder were identically structured in the proposed and Luo's models.

        \subsubsection{Hyperparameters}
            The other experimental conditions are identical to those used in \cite{Luo2019ISMIR}.
            The standard deviations of the prior distributions, $\sigyp$ and $\sigjt$, were constants: $\sigyp = e^{-2} \bm{1}_{\Dp}$ for all $\yp\in\mathcal{P}$ and $\sigjt = \bm{1}_{\Dt}$ for all $j\in\mathcal{T}$.
            The number of pitches and timbres were set as $|\mathcal{P}|=82$ and $|\mathcal{T}|=12$ in accordance with the numbers of pitches and musical instruments in the dataset, respectively.
            The batch size was set to 128.
            All methods were initialized by Xavier's initialization method, and the DNN parameters were optimized by an Adam optimizer with a learning rate of $10^{-4}$.
            The training was stopped when there was no improvement of the value of \eqref{eq:ELBO} without the second term about validation data for 1000 epochs.
		
    \subsection{Evaluation Metrics}
        To evaluate the efficiency of the proposed timbre embedding, we used the following two metrics.
        
        \subsubsection{Timbre Classification Accuracy}
                The first metric was the accuracy of predicting $\yt$ from the timbre latent variable.
                Since the posterior distribution $q(\yt|\X)$ is approximated as \eqref{eq:approx}, the \emph{maximum a posteriori} estimation of $\yt$ can be formulated as the maximization problem of $\mathbb{E}_{\zttil \sim q_{\phit}(\zttil | \X)}[p(\yt|\zttil)]$.
                By using the approximation 
                \begin{equation}
                    \mathbb{E}_{\zttil \sim q_{\phit}(\zttil | \X)}[p(\yt|\zttil)]\simeq p(\yt|\zttil = \xittil),
                \end{equation}
                we can predict $\yt$ as 
                \begin{equation}
                    \hatyt=\argmax_{j \in \mathcal{T}} p(\yt = j | \zttil = \xittil).
                \end{equation}
                The classification accuracy was computed as the ratio of the number of the correctly predicted timbre labels to that of the test data.

        \subsubsection{Hierarchical Separability}
            The second metric was for evaluating how close musical instruments belonging to the same instrument family are and how far those belonging to different instrument families are.
            The musical instruments in the dataset were grouped into the following four instrument families.
            \begin{itemize}
                \item Woodwind: English horn, saxophone, bassoon, clarinet, flute, and oboe (six instruments).
                \item Brass: French horn, tenor trombone, and trumpet (three instruments).
                \item Strings: Violin and violoncello (two instruments).
                \item Keyboard: Piano (one instrument).
            \end{itemize}
            If the timbre latent space captures the hierarchy of the instruments, $\mujtil$ and $\mujprimetil$ should be close to each other for the same instrument family and far from each other for a different instrument family.
            Hence, we define a metric (hierarchical separability) based on the distances in the timbre latent space as follows.
            
            The distance between $\mujtil$ and $\mujprimetil$ was computed in accordance with \eqref{eq:hyperbolic distance}.
            We call it the inter-instrument distance.
            Let $\mathcal{C}^{(\same)} \subset \mathcal{T} \times \mathcal{T}$ be the set of timbre index pairs belonging to the same instrument families and $\mathcal{C}^{(\diff)} \subset \mathcal{T} \times \mathcal{T}$ be that belonging to different instrument families.
            Note that $|\mathcal{C}^{(\same)}|=19$ and $|\mathcal{C}^{(\diff)}|=47$.
            The averages of inter-instrument distances of all $(j,j')\in\mathcal{C}^{(\same)}$ and $\mathcal{C}^{(\diff)}$ were respectively computed as
            \begin{align}
                \Dsame &= \frac{1}{|\mathcal{C}^{(\same)}|} \sum_{(j,j') \in \mathcal{C}^{(\same)}} \DHKDt(\mujtil, \mujprimetil), \\
                \Ddiff &= \frac{1}{|\mathcal{C}^{(\diff)}|} \sum_{(j,j') \in \mathcal{C}^{(\diff)}} \DHKDt(\mujtil, \mujprimetil).
            \end{align}
            Note that the distance $\DHKDt$ is replaced by the Euclidean distance in the evaluation of Luo's method.
            The hierarchical separability is given as
            \begin{equation}
                S = \frac{\Ddiff}{\Dsame}.
            \end{equation}
            A larger $S$ means that the instruments belonging to the same instrument family are embedded closer together and those belonging to different instrument families are embedded farther apart.
            
    \subsection{Results and Discussion}
        
        \begin{table}[t] 
            \centering{
                \caption{Timbre classification accuracies (\%) with various geometries, curvature radii, and dimensions of the timbre latent spaces} 
                \vspace{6pt}
                \label{tab: accuracy}
                \begin{tabular}{cc|cccc}
                    \toprule
                    \multirow{2}{*}{Geometry} & \multirow{2}{*}{$R$} & \multicolumn{4}{c}{$\Dt$}          \\
                                        &    &            2  &            4  &            8  &            16  \\
                    \midrule
                    Euclidean           &  - &         39.5  &         55.9  &         91.0  & \textbf{100.0} \\
                    \multirow{7}{*}{
                    	\begin{tabular}{c}
                    		Hyperbolic\\
                    		(Proposed)
                    	\end{tabular}
                    }                   &  1 &         47.5  &         35.6  &         44.1  &          96.0  \\
                                        &  2 &         35.6  &         65.0  &         88.7  &          98.9  \\
                                        &  5 &         42.9  &         62.1  &         93.8  &          98.3  \\
                                        & 10 &         50.3  &         68.9  &         85.3  &          98.3  \\ 
                                        & 20 &         40.1  &         78.5  &         87.6  &          98.3  \\
                                        & 50 &         49.7  &         74.6  & \textbf{96.6} &          99.4  \\
                                        &100 & \textbf{59.9} & \textbf{81.9} &         94.4  &          98.3  \\ 
                    \bottomrule
                \end{tabular}
            }
        \end{table}
        
        Table~\ref{tab: accuracy} shows timbre classification accuracies under each condition.
        With a larger $\Dt$, both the proposed and Luo's methods provided higher accuracies.
        As $\Dt$ decreased, the accuracy of Luo's method greatly decreased.
        By contrast, the accuracy of the proposed method remained higher than that of Luo's method.
        For example, when $\Dt=4$, the proposed method with $R=100$ provided 26.0 points higher accuracy than Luo's method.
        This result clearly shows that the hyperbolic space can efficiently represent the timbre space.
        
        For $\Dt=16$, the curvature radius $R$ did not greatly affect the accuracy of the proposed method.
        This is because $\Dt$ was sufficiently large to obtain high accuracy and $R$ affects the accuracy less than does $\Dt$.
        On the other hand, for $\Dt=2,4,$ and $8$, the larger $R$ tended to provide a higher accuracy.
        This result shows the importance of the curvature at a lower $\Dt$.
        The accuracy of the proposed method with $R=1$ was significantly lower than the other methods at $\Dt=4$ and $8$.
        With smaller curvature radii, we experimentally observed numerical instability in the computation of the exponential and logarithm maps, which may be one of the causes of the performance drop.

        \begin{table}[t] 
            \caption{Hierarchical separability $S$ with various geometries, curvature radii, and dimensions of the timbre latent spaces} 
            \vspace{6pt}
            \label{tab: ratio}
            \centering
                \begin{tabular}{cc|cccc}
                    \toprule
                    \multirow{2}{*}{Geometry} & \multirow{2}{*}{$R$} & \multicolumn{4}{c}{$\Dt$}  \\
                                        &    &             2  &             4  &             8  &             16  \\
                    \midrule
                    Euclidean           &  - & \textbf{2.149} &         1.332  &         1.132  &          1.071  \\
                    \multirow{7}{*}{
                    	\begin{tabular}{c}
                    		Hyperbolic\\
                    		(Proposed)
                    	\end{tabular}
                    }                   &  1 &         1.185  &         0.995  & \textbf{1.287} & \textbf{1.129} \\
                                        &  2 &         1.573  &         1.358  &         1.166  &         1.107  \\
                                        &  5 &         1.762  & \textbf{1.406} &         1.080  &         1.082  \\
                                        & 10 &         1.359  &         1.245  &         1.062  &         1.061  \\ 
                                        & 20 &         1.421  &         1.283  &         1.071  &         1.067  \\
                                        & 50 &         1.304  &         1.196  &         1.095  &         1.055  \\
                                        &100 &         1.369  &         1.247  &         1.215  &         1.077  \\ 
                    \bottomrule
                \end{tabular}
        \end{table}
        
        Table~\ref{tab: ratio} shows the hierarchical separability $S$ under each condition.
        The proposed method with adequate $R$ achieved a higher hierarchical separability except at $\Dt=2$.
        This result suggests that the use of the hyperbolic space can capture the hierarchical relationship in an unsupervised manner.
        At $\Dt=2$, the proposed method had a lower hierarchical separability than Luo's method.
        This may be alleviated by increasing the bits of precision because the lower-dimensional hyperbolic space requires more bits of precision for embedding tree-structured graphs \cite{Sala2018ICML}.
        We leave its further investigation as a future work.

\section{Conclusion}
We investigated the effect of introducing a hierarchy-inducing latent space for efficient timbre embeddings in VAE-based MISS.
The idea of the proposed method is to utilize the hierarchy of the musical instrument classification system for MISS.
For capturing the hierarchy, we extended Luo's method so that its latent space of timbre is a hyperbolic space.
In the proposed method, we used a pseudo-hyperbolic Gaussian distribution, a hyperbolic-space counterpart of a normal distribution.
It enables us to use a reparameterization trick in the hyperbolic space and train the proposed model by gradient descent algorithms commonly used in VAE training.
Results of experiments showed that the hyperbolic space can be used as an efficient latent space of timbre compared with the Euclidean space.

\end{document}